\documentclass[aps,nofootinbib,superscriptaddress,pra,twocolumn,,showpacs]{revtex4-2}
\usepackage{amsfonts, amsmath, amssymb, bm, bbm, graphicx, epsfig, epstopdf}
\usepackage{braket}
\usepackage{dcolumn}
\usepackage{textcomp}
\usepackage{placeins}
\usepackage[bottom]{footmisc}
\usepackage{color}
\usepackage{soul}
\usepackage[caption=false]{subfig} 
\definecolor{forestgreen}{RGB}{34,139,34}
\usepackage[normalem]{ulem} % Need for strikethrough
\usepackage[normalem]{ulem} % Need for strikethrough
\usepackage{xcolor}
\definecolor{girlishpink}{RGB}{211, 129, 195}
\usepackage[colorlinks=true, linkcolor=teal, citecolor=blue, urlcolor=girlishpink]{hyperref}
\usepackage[capitalise,noabbrev,nameinlink]{cleveref}
\usepackage{cleveref}

\crefname{figure}{Fig.}{Figs.}
\Crefname{figure}{Fig.}{Figs.}

\crefname{equation}{Eq.}{Eqs.}
\Crefname{equation}{Eq.}{Eqs.}

\begin{document}
\title{From quantum storage to amplification: the effect of unwanted couplings and an additional level in cavity-based ensemble quantum memories}

\author{Jia-Wei Ji}
\email{quantum.jiawei.ji@gmail.com}
\affiliation{Institute for Quantum Science and Technology, and Department of Physics \& Astronomy, University of Calgary, 2500 University Drive NW, Calgary, Alberta T2N 1N4, Canada}

\author{Christoph Simon}
\email{christoph.simon@gmail.com}
\affiliation{Institute for Quantum Science and Technology, and Department of Physics \& Astronomy, University of Calgary, 2500 University Drive NW, Calgary, Alberta T2N 1N4, Canada}

% \author{Faezeh Kimiaee Asadi\authormark{1,4}, Janish Kumar\authormark{2}, Jiawei Ji\authormark{1}, Khabat Heshami\authormark{3} and Christoph Simon\authormark{1}}

% \address{\authormark{1}Institute for Quantum Science and Technology, and Department of Physics \& Astronomy, University of Calgary, 2500 University Drive NW, Calgary, Alberta T2N 1N4, Canada\\
% \authormark{2}Department of Physics, Indian Institute of Technology Roorkee,
% Uttarakhand-247667, India\\
% \authormark{3}National Research Council of Canada, 100 Sussex Drive, Ottawa, Ontario K1A 0R6, Canada\\
% \authormark{4}The authors contributed equally to this work.}
%\authormark{*}Corresponding author: 

\begin{abstract}
Quantum-memory models often reduce complex level structures to an idealized $\Lambda$ system, potentially missing nearby levels and unwanted couplings that can qualitatively alter the predicted performance. Here, we study an extension of a cavity-based $\Lambda$-type ensemble memory, a four-level model with unwanted couplings from both the control field and signal, using a fully quantum treatment. We derive explicit expressions for the single-photon storage efficiency, retrieval efficiency, and fidelity, and on this basis identify three distinct dynamical regimes: stable, threshold, and unstable. Within the stable regime, we additionally discriminate between two qualitatively different sub-regimes. Applying the theory to warm-vapor-inspired parameters, we determine the conditions under which the system can still operate as a high-quality quantum memory. More generally, our results provide a practical framework for distinguishing genuine memory operation from amplification and for optimizing realistic quantum memories beyond idealized models.
\end{abstract}

\maketitle

%\def\thefootnote{*}\footnotetext{These authors contributed equally to this work}\def\thefootnote{\arabic{footnote}}
 
%\section{Introduction}\label{ssec:intro}
\section{Introduction}
\label{sec:intro}
Quantum memories play a central role in quantum technologies by enabling the storage and coherent processing of quantum information, with a broad impact on both quantum computing and quantum communication~\cite{lvovsky2009optical, bussieres2013prospective, heshami2016quantum}. For long-distance communication, they are indispensable in quantum repeater schemes, which mitigate channel loss and thereby enable scalable quantum networks and ultimately a quantum internet~\cite{kimble2008quantum, simon2017towards, Wehnereaam9288, RevModPhys.95.045006}.

A variety of absorptive quantum memory protocols have been developed, including electromagnetically induced transparency (EIT)~\cite{phillips2001storage,fleischhauer2005electromagnetically,chaneliere2005storage,longdell2005stopped, ma2022high}, Raman memories~\cite{michelberger2015interfacing,saunders2016cavity,guo2019high,heshami2014raman}, and schemes based on Autler-Townes splitting (ATS)~\cite{saglamyurek2018coherent,saglamyurek2019single}. Each offers distinct trade-offs, making different approaches advantageous across various physical platforms. Memory performance is commonly quantified by efficiency and fidelity~\cite{lei2023quantum,simon2010quantum}. The fidelity is typically defined by the signal-to-noise ratio (SNR). High efficiency is often limited by intrinsic loss channels in the system, including decoherence and decay. Fidelity can be reduced by loss and added noise due to unwanted couplings and imperfect control~\cite{lauk2013fidelity, saunders2016cavity, nunn2017theory}. Depending on the application, the memory efficiency and fidelity requirements could vary, but improving either generally enhances system-level performance.

Although simultaneously achieving high efficiency and low noise in absorptive quantum memories remains nontrivial~\cite{lei2023quantum}, especially in traditional $\Lambda$-type platforms, important progress has been made on understanding unwanted noise processes in effective $\Lambda$-type models, including four-wave mixing and its suppression~\cite{PhysRevA.90.033823,nunn2017theory,thomas2019raman}. Meanwhile, a growing body of experiments has demonstrated high-performance quantum-memory operation in warm-vapor and cavity-enhanced platforms~\cite{Hosseini2011,reim2011single,saunders2016cavity,guo2019high,ma2022high,Namazi2017,wu2025ai}. The key open question is therefore not whether good memory performance can be achieved, but how to characterize the conditions under which it remains robust in realistic multilevel systems. Because simplified $\Lambda$-type models do not explicitly include nearby auxiliary levels and the additional off-resonant couplings induced by the control and signal fields, they can miss important noise and gain channels. Previous work on NV-center memories has captured amplification arising from such undesired couplings using numerical and semiclassical methods~\cite{asadiunwanted}, but a general, fully quantum treatment has remained unavailable.

%Although several solutions have been proposed to mitigate noise from unwanted coupling in $\Lambda$-type memories~\cite {PhysRevA.90.033823,nunn2017theory,thomas2019raman}, they typically ignore couplings to other levels. This simplified treatment could lead to inaccurate predictions of memory performance. A systematic understanding of these imperfections remains largely lacking. Developing ways to quantify and mitigate them is therefore essential for optimizing quantum-memory performance. Previous work has focused on quantifying the signal amplification induced by undesired couplings in NV centers using numerical and semi-classical methods~\cite{asadiunwanted}, but lacks a full quantum treatment.        

In this work, we study a four-level extension of a cavity-based $\Lambda$-type ensemble memory with unwanted couplings from both the control and signal fields. In realistic media, such unwanted couplings may arise either through a distinct nearby auxiliary excited state or through the same excited state as the desired transition, depending on the level structure and field polarizations. Here we focus on the former situation and introduce a separate excited state \(\ket{e'}\), which provides the simplest explicit four-level model of the leading gain and noise channels beyond the idealized $\Lambda$ description. Although realistic media generally contain more than four levels, this model remains analytically tractable while capturing the dominant nearby-level corrections. Using a fully quantum approach, we derive analytical expressions for the storage efficiency, retrieval efficiency, and operational fidelity, and identify three dynamical regimes: stable, threshold, and unstable. We further resolve two qualitatively distinct sub-regimes within the stable sector. Bounded, below-threshold dynamics does not by itself guarantee gain-free memory operation: sufficiently weak unwanted couplings still allow high-quality storage and retrieval, whereas stronger couplings can already produce finite amplification below threshold. At the threshold and in the unstable regime, the output grows linearly and exponentially in time, respectively, signaling a crossover from memory behavior to amplification. We finally apply the theory to warm-vapor-inspired parameters and determine when such a four-level system can still operate as a high-performance quantum memory.

%In this work, we consider a four-level system in the cavity with unwanted couplings from both the control field and the signal. We calculate the number of retrieved photons in two regimes: the memory regime and the source regime. In the memory regime, the finite efficiency and fidelity of this generic memory are provided in the presence of these couplings using a full quantum approach. However, in the source regime, the number of noise photons grows exponentially, leading to an inoperable memory. We further apply this approach to warm vapors and specify the experimental conditions under which this system operates as a good memory.   

%\textcolor{red}{could be generalized to more levels? Does the same treatment also apply here? And the rationale for considering this scheme is that it's a minimal level structure that gives rise to the issue and this is an extension of our previous work.}

The remainder of the paper is organized as follows. Section~\ref{sec:model} introduces the four-level model, derives the effective spin-wave dynamics, and states the approximations that control the analysis. Section~\ref{sec:retrieval} treats retrieval, beginning with the bounded stable regime and then turning to the threshold and unstable cases. Section~\ref{sec:storage} develops the corresponding storage problem and shows explicitly how unwanted couplings break the usual time-reversal relation between storage and retrieval processes. Section~\ref{sec:fidelity} presents the total memory efficiency and the corresponding fidelity for the stable, threshold, and unstable regimes, and discusses the differences among these regimes. Finally, Sec.~\ref{sec:conclusion} summarizes the main conclusions and outlook.

\section{Four-Level Model and Effective Dynamics}
\label{sec:model}
\begin{figure}
\centering
  \includegraphics[width=0.8\linewidth]{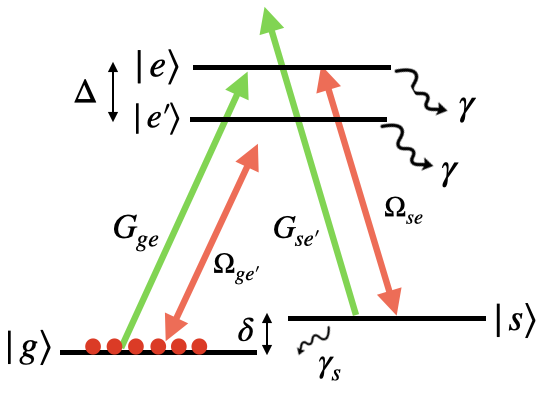}
  \caption{
The memory level diagram. The input signal drives the desired \(\ket{g}-\ket{e}\) transition with coupling strength \(G_{ge}\), while the control field drives the desired \(\ket{s}-\ket{e}\) transition with time-dependent Rabi frequency \(\Omega_{se}(t)\). The unwanted couplings are modeled explicitly through a distinct auxiliary excited state \(\ket{e'}\): the control field off-resonantly couples \(\ket{g}-\ket{e'}\) with strength \(\Omega_{ge'}(t)\), and the signal off-resonantly couples \(\ket{s}-\ket{e'}\) with strength \(G_{se'}\). This four-level scheme should be viewed as an extension of the idealized \(\Lambda\)-type memory, capturing the leading nearby-level correction while remaining analytically tractable. Both excited states \(\ket{e}\) and \(\ket{e'}\) are assumed to have the same decoherence rate \(\gamma\), the spin state \(\ket{s}\) has decoherence rate \(\gamma_s\), the ground-state splitting is \(\delta\), and the excited-state splitting is \(\Delta\).
}
  %\caption{The memory level diagram. The $\ket{g}$-the input signal couples $\ket{e}$ transition with strength $G_{ge}$. The control field is coupled to the $\ket{e}$- $\ket{s}$ transition via a time-dependent Rabi frequency, $\Omega_{se}(t)$. The control field and the signal also off-resonantly couple to the $\ket{g}$-$\ket{e'}$ transition with the strength $\Omega'_{ge'}(t)$ and the $\ket{s}$-$\ket{e'}$ transition with the strength $G_{se'}$ respectively. Both the excited states $\ket{e}$ and $\ket{e'}$ are assumed to have the same decoherence rate $\gamma$, and the spin state $\ket{s}$ has a decoherence rate of $\gamma_s$. The ground state splitting between the states $\ket{g}$ and $\ket{s}$ is $\delta$, and the frequency difference between the two excited states is $\Delta$.}
  \label{levels}
  \end{figure}

We now introduce the minimal four-level model shown in Fig.~\ref{levels}. The two excited states $\ket{e}$ and $\ket{e'}$ constitute the simplest explicit level structure that captures the unwanted-coupling physics of interest and directly extend the setting of Ref.~\cite{asadiunwanted}. Both excited states are taken to have the same decoherence rate $\gamma$. The atoms are prepared initially in the ground state $\ket{g}$, while the spin state $\ket{s}$ has decoherence rate $\gamma_s$. The input signal addresses the transitions $\ket{g}-\ket{e}$ and $\ket{s}-\ket{e'}$ with strengths $G_{ge}$ and $G_{se'}$, respectively. The control field drives $\ket{s}-\ket{e}$ with Rabi frequency $\Omega_{se}$ and also off-resonantly couples $\ket{g}-\ket{e'}$ with strength $\Omega_{ge'}$.

Given that this memory interacts with the input signal and control field, we can write a set of Heisenberg-Langevin equations to characterize evolution~\cite{asadiunwanted}:

\begin{equation}
\begin{aligned}
\dot{\hat{\sigma}}_{ge'} ={}&
-(\gamma+i\Delta)\hat{\sigma}_{ge'}
+i\hat{a}G_{se'}\hat{\sigma}_{gs}e^{i\delta t}\\
&+i\Omega_{ge'}N e^{-i\delta t}
+\hat{F}_{ge'}, \\
\dot{\hat{\sigma}}_{ge} ={}&
-\gamma \hat{\sigma}_{ge}
+i\hat{a}G_{ge}N
+i\Omega_{se}\hat{\sigma}_{gs}
+\hat{F}_{ge}, \\
\dot{\hat{\sigma}}_{gs} ={}&
-\gamma_s \hat{\sigma}_{gs}
-i\hat{a}G_{ge}\hat{\sigma}_{ge}^{\dagger}
+i\Omega_{se}^{*}\hat{\sigma}_{ge} \\
&+i\hat{a}^{\dagger}G_{se'}^{*}\hat{\sigma}_{g'e'}e^{-i\delta t}
+\hat{F}_{gs}, \\
\dot{\hat{\sigma}}_{se} ={}&
-\gamma \hat{\sigma}_{se}
+i\hat{a}G_{ge}\hat{\sigma}_{gs}^{\dagger}
+\hat{F}_{se}, \\
\dot{\hat{a}} ={}&
-\kappa \hat{a}
+\sqrt{2\kappa}\,\hat{a}_{\mathrm{in}}
+iG_{ge}^{*}\hat{\sigma}_{ge}.
\end{aligned}
\label{eq:dynamics}
\end{equation}
where $\hat{\sigma}_{ab}$ with $a=g,s$ and $b=s,e,e'$ represents the annihilation operator for the corresponding transition. We assume that almost all atoms stay in the ground state during evolution such that $\hat{\sigma}_{gg}\approx N,\hat{\sigma}_{ee}\approx\hat{\sigma}_{e'e'}\approx\hat{\sigma}_{ss}\approx\hat{\sigma}_{se'}\approx0$~\cite{gorshkov2007photon}. $N$ is the number of atoms in the ground state. In addition, we keep $\hat{\sigma}_{se}$ as the control field drives its evolution.

Now, we first ignore the Langevin noise operators in Eq.~\eqref{eq:dynamics} as all atoms are prepared in the ground state, in which case these normally ordered operators do not contribute to noise~\cite{gorshkov2007photon}. To further simplify this set of equations, we adopt the assumptions from the previous study~\cite{asadiunwanted} that both the excited states $\ket{e}$ and $\ket{e'}$ can be eliminated adiabatically. This assumption is justified because we operate in the electromagnetically induced transparency (EIT) regime for the desired couplings, while the undesired couplings to the auxiliary excited state $\ket{e'}$ are far off resonance and can therefore be neglected to a good approximation. Furthermore, we assume that the system is in the bad-cavity regime. A sufficient set of order-of-magnitude conditions for adiabatic elimination of modes in Eq.~\eqref{eq:dynamics} is
\begin{equation}
\begin{aligned}
\sqrt{\gamma^2+\Delta^2} &\gg |\Omega_{ge'}|,\ \sqrt{N}|G_{se'}|,\ \tau^{-1},\\
\gamma &\gg |\Omega_{se}|,\ \sqrt{N}|G_{ge}|,\ \tau^{-1},\\
\kappa &\gg \sqrt{N}|G_{ge}|,\ \tau^{-1},
\end{aligned}
\end{equation}
where \(\tau\) denotes the characteristic timescale of the input pulse.

Now, Setting $\dot{\hat{\sigma}}_{ge'}=\dot{\hat{\sigma}}_{ge}=\dot{\hat{\sigma}}_{se}=\dot{\hat{a}}=0$, we can rewrite the equations in Eq.~(\ref{eq:dynamics}) as:

\begin{equation}
\begin{aligned}
\dot{\hat{\sigma}}_{gs}=&-(\gamma_s+\frac{|\Omega_{se}|^2}{\gamma})\hat{\sigma}_{gs}-\frac{|G_{se'}|^2}{\gamma+i\Delta}\frac{\hat{\beta}^\dagger\hat{\beta}}{\alpha^2}\hat{\sigma}_{gs}\\
&-\frac{N\Omega^*_{eg}G_{ge}}{\gamma}\frac{\hat{\beta}}{\alpha}
-\frac{NG^*_{se'}\Omega_{ge'}}{\gamma+i\Delta}\frac{\hat{\beta}^\dagger}{\alpha}e^{-2i\delta t}\\
&-\frac{|G_{ge}|^2}{\gamma}\frac{\hat{\beta}\hat{\beta}^\dagger}{\alpha^2}\hat{\sigma}^\dagger_{gs},
\end{aligned}
\label{eq:adibatic}
\end{equation} 
where $\hat{\beta}=\sqrt{2\kappa}\gamma\hat{a}_{\text{in}}-\Omega_{se}G^*_{ge}\hat{\sigma}_{gs}$ and $\alpha=\gamma\kappa+|G_{ge}|^2N$. This equation comprises both linear terms and nonlinear terms. The nonlinear contributions are cubic, involving products of three operators, namely $\hat{\beta}$, $\hat{\beta}^\dagger$, and $\hat{\sigma}_{gs}$. Nonetheless, a more detailed numerical analysis shows that these nonlinear terms are negligible relative to the linear contributions, as quantitatively demonstrated in \cite{asadiunwanted}. In Appendix~\ref{appa}, we mathematically justify that the non-linear terms can indeed be dropped in the weak-excitation regime. 

%\subsection{Low-excitation justification}
%\label{sec:detailed-approx}

%The derivation of Eq.~\eqref{eq:adibatic} already assumes a low-excitation regime, namely that almost all atoms remain in the ground state, so that the spin-wave excitation is weak. We therefore introduce a small parameter \(\varepsilon\) and count
%\begin{equation}
%\hat{a}_{\mathrm{in}}=O(\varepsilon),\qquad
%\hat{\sigma}_{gs}=O(\varepsilon),\qquad
%\hat{\sigma}_{g's}=O(\varepsilon).
%\label{A3}
%\end{equation}
%Since \(\hat{\beta}\) is linear in \(\hat{a}_{\mathrm{in}}\) and \(\hat{\sigma}_{gs}\), it follows that
%\begin{equation}
%\hat{\beta}=O(\varepsilon),
%\qquad
%\hat{\beta}^\dagger\hat{\beta}=O(\varepsilon^2).
%\label{A4}
%\end{equation}
%Hence the linear terms in scale as
%\begin{equation}
%\left(\gamma_s+\frac{|\Omega_{se}|^2}{\gamma}\right)\hat{\sigma}_{gs}
%=O(\varepsilon),
%\quad
%\frac{\hat{\beta}}{\alpha}=O(\varepsilon),
%\quad
%\frac{\hat{\beta}^\dagger}{\alpha}=O(\varepsilon),
%\label{A5}
%\end{equation}
%whereas the non-linear terms scale as
%\begin{equation}
%\frac{\hat{\beta}^\dagger\hat{\beta}}{\alpha^2}\hat{\sigma}_{gs}
%=O(\varepsilon^3),
%\quad
%\frac{\hat{\beta}\hat{\beta}^\dagger}{\alpha^2}\hat{\sigma}_{gs}^\dagger
%=O(\varepsilon^3).
%\label{A6}
%\end{equation}
%Therefore, the non-linear contributions are smaller than the retained ones by a factor \(O(\varepsilon^2)\), and Eq.~\eqref{eq:simp} is simply the first-order truncation of Eq.~\eqref{eq:adibatic} in the weak-excitation expansion.

By dropping the non-linear terms in Eq. (\ref{eq:adibatic}), we obtain:

\begin{equation}
\begin{aligned}
\dot{\hat{\sigma}}_{gs}=&-\Gamma_{\text{eff}}\hat{\sigma}_{gs}+\frac{e^{-2i\delta t}G^*_{se'}\Omega_{ge'}G_{ge}\Omega^*_{se}N}{(\gamma+i\Delta)\alpha}\hat{\sigma}^\dagger_{gs}\\
&-\frac{N\sqrt{2\kappa}\Omega^*_{se}G_{ge}}{\alpha}\hat{a}_{\text{in}}-\frac{e^{-2i\delta t}N\sqrt{2\kappa}\gamma G^*_{se'}\Omega_{ge'}}{(\gamma+i\Delta)\alpha}\hat{a}^\dagger_{\text{in}},
\end{aligned}
\label{eq:simp}
\end{equation}
where $\Gamma_{\text{eff}}=\gamma_s+|\Omega_{se}|^2/\gamma-N|\Omega_{se}|^2|G_{ge}|^2/\gamma\alpha$ is the effective decay rate for the mode $\hat{\sigma}_{gs}$. This decay rate can also be rewritten as $\Gamma_{\text{eff}}=\gamma_s+|\Omega_{se}|^2/[\gamma(C+1)]$ with $C=G^2_{ge}N/\gamma\kappa$ being the cooperativity. In the majority of practical scenarios, the spin decoherence rate can be safely neglected, as it is significantly smaller than the control-field-induced transition rates, that is, $\gamma_s \ll |\Omega_{se}|^2 / [\gamma (C+1)]$.

\section{Retrieval Dynamics}
\label{sec:retrieval}

This section is devoted to the retrieval process. We derive the optimal retrieval efficiency for three regimes: stable, threshold, and unstable, and discuss the conditions for each regime.

\subsection{Retrieval efficiency in the stable regime}
\label{sec:detailed-ret-stable}

The evolution of the collective spin mode is not only driven by the input field but also by the unwanted couplings $\Omega_{ge'}$ and $G_{se'}$. In the retrieval process, since there is no signal field, we can write Eq. (\ref{eq:simp}) as

\begin{equation}
\dot{\hat{\sigma}}_{gs}=-\Gamma_{\text{eff}}\hat{\sigma}_{gs}+\frac{e^{-2i\delta t}G^*_{se'}\Omega_{ge'}G_{ge}\Omega^*_{se}N}{(\gamma+i\Delta)\alpha}\hat{\sigma}^\dagger_{gs}.
\label{eq:retrievaldynamics}
\end{equation}
In the low-excitation regime, we consider a collective spin mode between the ground state $\ket{g}$
and a metastable state $\ket{s}$ as
\begin{equation}
    \hat\sigma_{gs} = \sqrt{N}\,\hat b,
\end{equation}
with $N$ atoms. The effective retrieval
dynamics of $\hat b$ is given by
\begin{equation}
\dot{\hat b}=-\Gamma_{\mathrm{eff}}\,\hat b+ g_{\text{eff}}\,e^{-2i\delta t}\,\hat b^\dagger,
\label{eq:b-eom}
\end{equation}
with effective coupling:
\begin{equation}
g_{\text{eff}}=\Bigg|\frac{G_{se'}^* \Omega_{ge'} G_{ge} \Omega_{se}^* N}{(\gamma + i\Delta)\,\alpha}\Bigg|.
\end{equation}
In general, the effective coupling $g_{\text{eff}}$ is complex, $g_{\text{eff}}=|g_{\text{eff}}|e^{i\phi_g}$, reflecting the phases of the control and cavity fields. However, the retrieval dynamics depends on $g_{\text{eff}}$ only through $g_{\text{eff}}$ and $g^*_{\text{eff}}$ in a symmetric way. We can choose a certain phase so that the effective coupling is real and non-negative.

%%%%%%%%%%%%%%%%%%%%%%%%%%%%%%%%%%%%%%%%%%%%%%%%%%%%%%%%%%%%

Introduce the rotating frame
\begin{equation}
    \hat c(t) = \hat b(t)e^{i\delta t},
\end{equation}
which turns Eq.~(\ref{eq:b-eom}) into
\begin{equation}
    \dot{\hat c}
    =
    (-\Gamma_{\mathrm{eff}}+i\delta)\,\hat c
    + g_{\text{eff}}\,\hat c^\dagger.
    \label{eq:c-eom}
\end{equation}
The eigenvalues of the drift matrix are
\begin{equation}
    \lambda_\pm = -\Gamma_{\mathrm{eff}} \pm \mu,
    \qquad
    \mu \equiv \sqrt{g_{\text{eff}}^2 - \delta^2}.
\end{equation}
To ensure stability, the largest real part of the eigenvalues should be negative, i.e., $\lambda_+<0$. Thus, we need to work in the stable regime
\begin{equation}
    \Gamma_{\mathrm{eff}}^2 + \delta^2 - g^2_{\text{eff}} > 0,
    \label{eq:stability}
\end{equation}

The stable regime naturally separates into two sub-regimes:
\begin{align}
&\text{(i)}\quad g_{\mathrm{eff}}\ge \delta,
\quad
g_{\mathrm{eff}}^{2}<\Gamma_{\mathrm{eff}}^{2}+\delta^{2},
\label{eq:stable-realmu}\\
&\text{(ii)}\quad g_{\mathrm{eff}}<\delta.
\label{eq:stable-imagmu}
\end{align}
In sub-regime (i), \(\mu=\sqrt{g_{\mathrm{eff}}^{2}-\delta^{2}}\) is real and satisfies
\(0\le \mu<\Gamma_{\mathrm{eff}}\). In sub-regime (ii), \(\mu\) is purely imaginary,
\(\mu=i\mu'\) with \(\mu'=\sqrt{\delta^{2}-g_{\mathrm{eff}}^{2}}\), and the system is always stable.

The Heisenberg solution is a Bogoliubov transform
\begin{equation}
    \hat c(t) = u(t)\hat c(0)+v(t)\hat c^\dagger(0),
\end{equation}
with
\begin{align}
    u(t)
    &=
    e^{-\Gamma_{\mathrm{eff}}t}
    \biggl[
        \cosh(\mu t) + i\frac{\delta}{\mu}\sinh(\mu t)
    \biggr],
    \\
    v(t)
    &=
    e^{-\Gamma_{\mathrm{eff}}t}\frac{g_{\text{eff}}}{\mu}\sinh(\mu t).
\end{align}
Since $\hat b$ and $\hat c$ differ only by phase,
$\langle \hat b^\dagger\hat b\rangle = \langle \hat c^\dagger\hat c\rangle$.

We consider retrieval from a single collective spin excitation,
\begin{equation}
    \ket{\psi_0} = \ket{1_b} = \ket{1_c}.
\end{equation}
In fact, the retrieval efficiency is not contingent upon the single-excitation assumption, since it is formally defined as the ratio between the number of retrieved photons and the number of spin excitations.  

The spin population resulting from this excitation is
\begin{equation}
    n(t)
    \equiv
    \big\langle \hat b^\dagger(t)\hat b(t)\big\rangle_{\mathrm{sig}}
    =
    |u(t)|^2 + 2|v(t)|^2,
\end{equation}
which evaluates to
\begin{equation}
    n(t)
    =
    \frac{e^{-2\Gamma_{\mathrm{eff}} t}}{2\mu^2}
    \Big[
        3g_{\text{eff}}^2\cosh(2\mu t)
        -\bigl(g^2_{\text{eff}}+2\delta^2\bigr)
    \Big].
    \label{eq:n-of-t}
\end{equation}
In the stable regime, the integral over the entire
retrieval window converges:
\begin{equation}
    \int_0^\infty n(t)\,dt
    =
    \frac{2\Gamma_{\mathrm{eff}}^2 + 2\delta^2 + g_{\text{eff}}^2}
         {4\Gamma_{\mathrm{eff}}
          \bigl(\Gamma_{\mathrm{eff}}^2+\delta^2-g_{\text{eff}}^2\bigr)}.
    \label{eq:int-n}
\end{equation}
This is valid for the two sub-regimes: $g_{\mathrm{eff}}\geq\delta$ and $g_{\mathrm{eff}}<\delta$. 

%%%%%%%%%%%%%%%%%%%%%%%%%%%%%%%%%%%%%%%%%%%%%%%%%%%%%%%%%%%%

\begin{figure}
\centering
\includegraphics[scale=0.25]{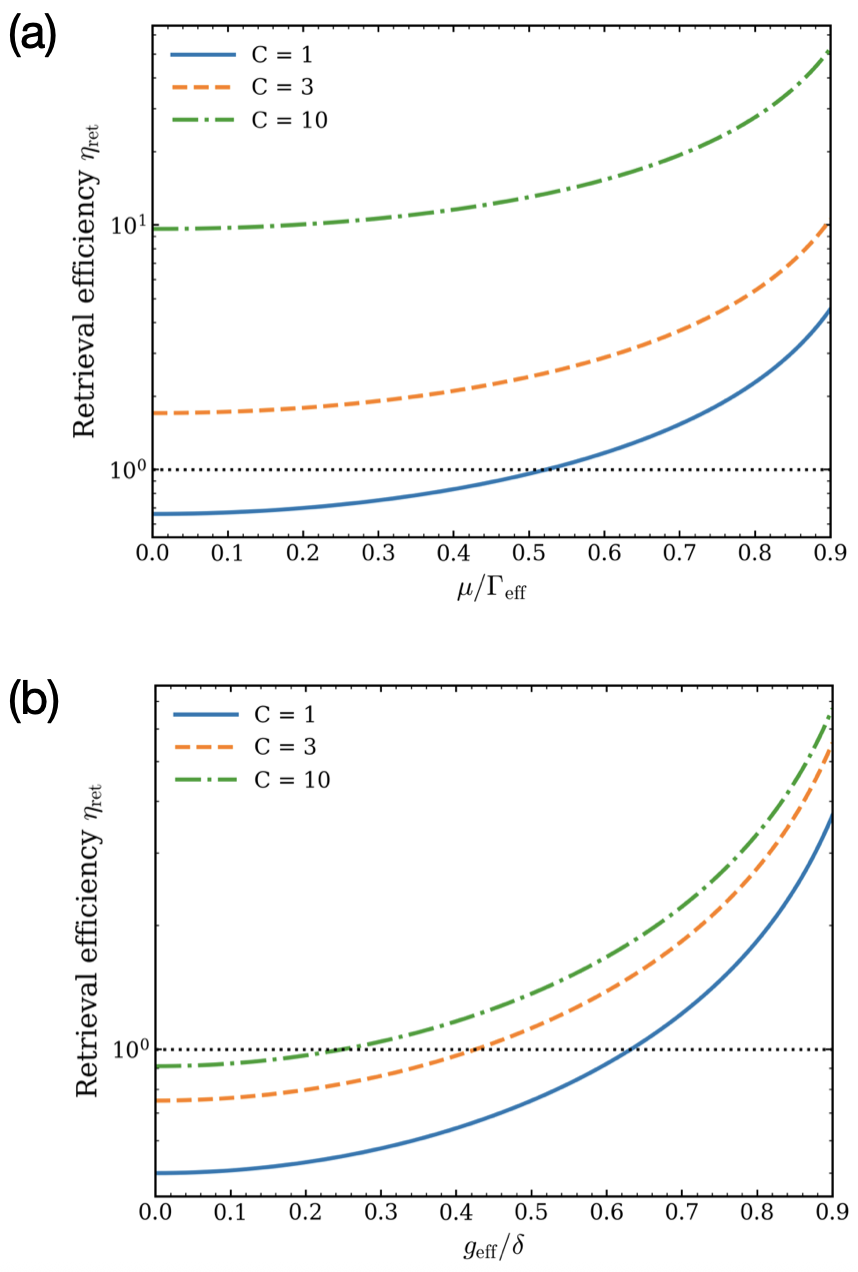}
\caption{
Retrieval efficiency \(\eta_{\mathrm{ret}}\) in (a), plotted for the stable regime
with \(g_{\mathrm{eff}}>\delta\) as a function of \(\mu/\Gamma_{\mathrm{eff}}\), where
\(\mu=\sqrt{g_{\mathrm{eff}}^2-\delta^2}\) is the real instability parameter of the effective spin-mode dynamics with $g_{\mathrm{eff}}$ the effective unwanted coupling, and $\delta$ the ground-state splitting. \(\Gamma_{\mathrm{eff}}=|\Omega_{se}|^2/[\gamma(C+1)]\) is the effective decay rate of the collective spin mode. Retrieval efficiency \(\eta_{\mathrm{ret}}\) in (b), plotted for the stable regime
with \(g_{\mathrm{eff}}<\delta\) as a function of \(g_{\mathrm{eff}}/\delta\), characterized by \(\Gamma_{\mathrm{eff}} > \mu'\) and \(\delta > g_{\mathrm{eff}}\), where \(\mu' = \sqrt{\delta^2 - g_{\mathrm{eff}}^2}\). For this panel, since in this case $\delta\gg\Gamma_{\mathrm{eff}}$, the performance is dictated by the ratio of $g_{\mathrm{eff}}$ to $\delta$. The curves correspond to cooperativities
\(C=|G_{ge}|^2N/(\gamma\kappa)=1,3,10\) with $G_{ge}$ the input signal coupling strength to the $\ket{g}-\ket{e}$ transition, $N$ the number of atoms in the ground state, $\gamma$ the decoherence rate of the excited state, and $\kappa$ the cavity decay rate.  
The fixed parameters are
\(\gamma=5.75~\mathrm{MHz}\),
\(\Omega_{se}=5~\mathrm{MHz}\),
\(\delta=1~\mathrm{MHz}\) for (a) and \(\delta=6.8~\mathrm{GHz}\) for (b), 
\(G_{ge}=20~\mathrm{kHz}\), 
\(\kappa=1~\mathrm{GHz}\), and
\(\Delta=0.816~\mathrm{GHz}\).
The horizontal dotted line in both panels marks \(\eta_{\mathrm{ret}}=1\), above which
the total output exceeds the ideal single-photon limit and indicates amplification.
}
\label{retrieval stable}
\end{figure}

During retrieval, \(\hat a_{\rm in}=0\), we have
\begin{equation}
0=-\gamma \hat\sigma_{ge}+iG_{ge}N\hat a+i\Omega_{se}\hat\sigma_{gs},
\quad
0=-\kappa \hat a+iG_{ge}^*\hat\sigma_{ge}.
\end{equation}
Solving these two equations yields
\begin{equation}
\hat a
=
-\frac{G_{ge}^*\Omega_{se}}{\alpha}\,\hat\sigma_{gs},
\quad
\alpha=\gamma\kappa+|G_{ge}|^2N.
\end{equation}
The cavity output field is then $\hat a_{\rm out}=\sqrt{2\kappa}\,\hat a$. Therefore, the photon flux into the desired output mode is
\begin{equation}
\Phi_{\rm sig}(t)
=
\langle \hat a_{\rm out}^\dagger \hat a_{\rm out}\rangle
=
2\kappa \frac{|G_{ge}|^2|\Omega_{se}|^2}{|\alpha|^2}
\langle \hat\sigma_{gs}^\dagger \hat\sigma_{gs}\rangle.
\end{equation}
Using \(\hat\sigma_{gs}=\sqrt{N}\hat b\), this becomes
\begin{equation}
\Phi_{\rm sig}(t)=2\Gamma_c\,n(t),
\quad
\Gamma_c\equiv
\frac{\kappa N |G_{ge}|^2 |\Omega_{se}|^2}{|\alpha|^2},
\end{equation}
where \(n(t)=\langle \hat b^\dagger(t)\hat b(t)\rangle\). With \(C=|G_{ge}|^2N/(\gamma\kappa)\) and
\begin{equation}
\Gamma_{\rm eff}
=
\gamma_s+\frac{|\Omega_{se}|^2}{\gamma(C+1)}
=
\gamma_s+\frac{\kappa |\Omega_{se}|^2}{\alpha},
\end{equation}
one finds
\begin{equation}
\Gamma_c
=
\frac{C}{C+1}\,\frac{\kappa |\Omega_{se}|^2}{\alpha}
=
\frac{C}{C+1}\left(\Gamma_{\rm eff}-\gamma_s\right).
\end{equation}
In practice, $\gamma_s\ll\Gamma_{\mathrm{eff}}$, this reduces to
\begin{equation}
\Gamma_c=\Gamma_{\rm eff}\frac{C}{C+1}.
\label{eq:gammac}
\end{equation}

For $g_{\text{eff}}\neq 0$ (still in the stable regime), the retrieval efficiency for one initial spin excitation is
\begin{equation}
    \eta_{\mathrm{ret}}
    =
    \int_0^\infty \Phi_{\mathrm{sig}}(t)\,dt
    =
    2\Gamma_c\int_0^\infty n(t)\,dt.
\end{equation}
Using Eq.~\eqref{eq:int-n} and Eq.~(\ref{eq:gammac}), we obtain
\begin{align}
    \eta_{\mathrm{ret}}
    &=
    2\Gamma_{\mathrm{eff}}\frac{C}{1+C}\,
    \frac{2\Gamma_{\mathrm{eff}}^2 + 2\delta^2 + g^2_{\text{eff}}}
         {4\Gamma_{\mathrm{eff}}
          \bigl(\Gamma_{\mathrm{eff}}^2+\delta^2-g^2_{\text{eff}}\bigr)}
    \nonumber\\[1ex]
    &=
    \frac{C}{1+C}\,
    \frac{2\Gamma_{\mathrm{eff}}^2 + 2\delta^2 + g^2_{\text{eff}}}
         {2\bigl(\Gamma_{\mathrm{eff}}^2+\delta^2-g^2_{\text{eff}}\bigr)}.
\end{align}
A compact form is
\begin{equation}
\eta_{\mathrm{ret}}=\frac{C}{1+C}\left[1+\frac{3g^2_{\text{eff}}}{2\bigl(\Gamma_{\mathrm{eff}}^2+\delta^2-g^2_{\text{eff}}\bigr)}\right].
\label{eq:retrieval}
\end{equation}
For $g_{\text{eff}} = 0$, this expression simplifies to $\eta_{\mathrm{ret}} = C/(1 + C)$, which is the expected result in the ideal case~\cite{gorshkov2007photon}. In the limit $g_{\text{eff}}^{2} \to \Gamma_{\mathrm{eff}}^{2} + \delta^{2}$ from below, the term in square brackets diverges, indicating that the system approaches the threshold for instability.

\begin{figure}
\centering
\includegraphics[scale=0.45]{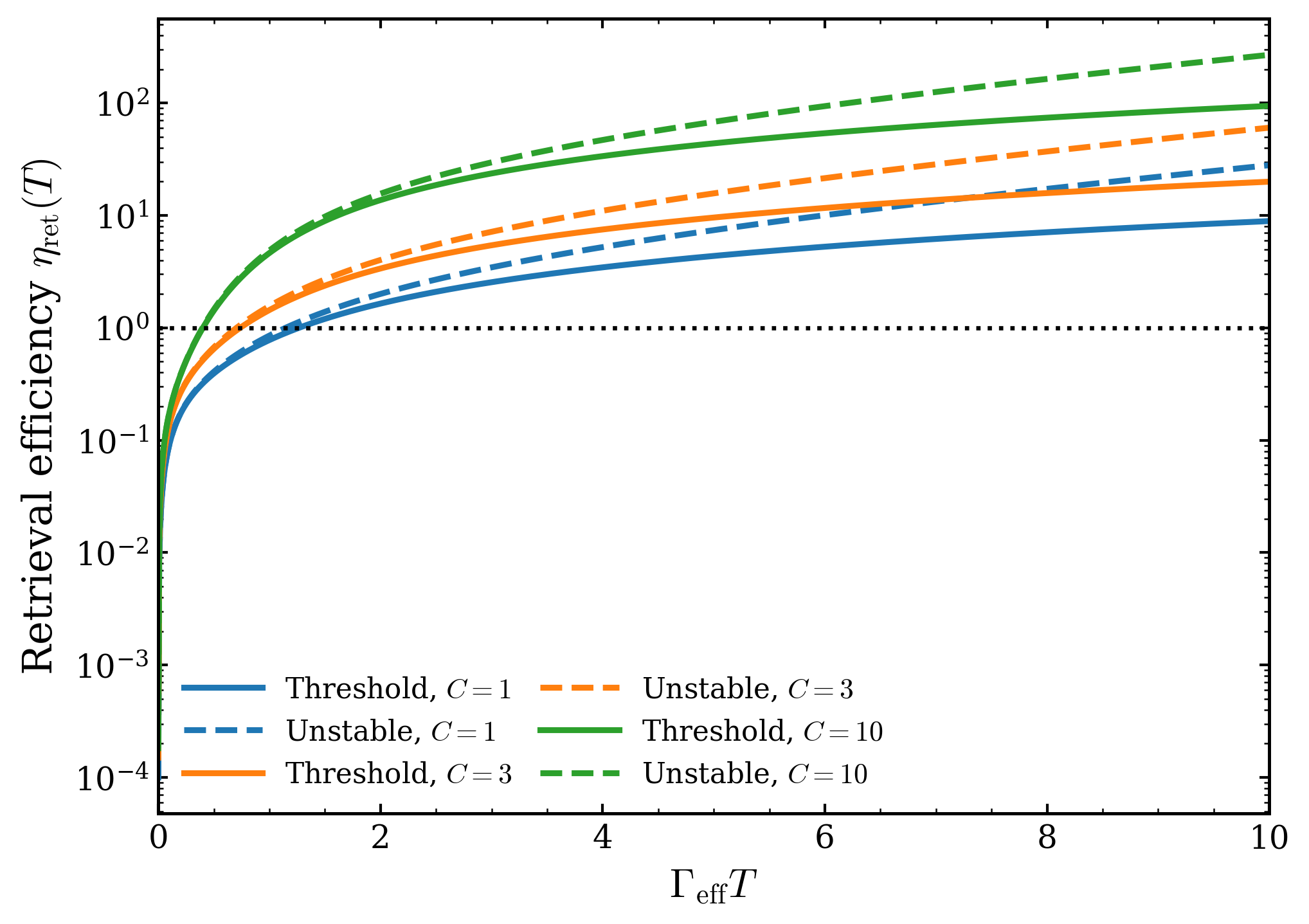}
\caption{
Retrieval efficiency for the threshold and unstable regimes, plotted as a function of the dimensionless total protocol time $\Gamma_{\mathrm{eff}}T$. The threshold regime satisfies $\Gamma_{\mathrm{eff}}=\mu$, and the unstable regime satisfies $\Gamma_{\mathrm{eff}}<\mu$. The curves correspond to cooperativities
\(C=|G_{ge}|^2N/(\gamma\kappa)=1,3,10\) in each regime. The fixed parameters are the same as those in Fig.~\ref{retrieval stable} with $\delta=1$ MHz. The horizontal dotted line marks \(\eta_{\mathrm{tot}}=1\), above which
the total output exceeds the ideal single-photon limit and indicates amplification. In these two cases, the memory effectively operates as a photon source, emitting photons continuously over time.}
\label{retrieval unstable}
\end{figure}

The previous results were derived under the simplifying assumption that the control fields are time-independent. However, if the control field \(\Omega_{se}(t)\) acquires an explicit time dependence while all effective rates share the same envelope, we can write
\begin{equation}
\Gamma_{\mathrm{eff}}(t)=f(t)\Gamma_0,\; g_{\text{eff}}(t)=f(t)g_0,\; \Gamma_c(t)=f(t)\Gamma_{c0},
\end{equation}
with $f(t)\propto|\Omega_{se}(t)|^2$, and
$\Gamma_{c0}=\Gamma_0\,C/(1+C)$. Introducing the ``control clock''
\begin{equation}
    \tau(t) = \int_0^t f(t')\,dt',
\end{equation}
the equation of motion becomes
\begin{equation}
    \frac{d\hat c}{d\tau}
    =
    (-\Gamma_0+i\delta)\hat c + g_0\hat c^\dagger,
\end{equation}
identical in form to Eq.~(\ref{eq:c-eom}) with $\Gamma_{\mathrm{eff}}
\to\Gamma_0$ and $g_{\text{eff}}\to g_0$. The photon flux is
\begin{equation}
    \Phi_{\mathrm{sig}}(t)
    =
    2\Gamma_c(t)\,n(t)
    =
    2f(t)\,\Gamma_{c0}\,n\bigl(\tau(t)\bigr),
\end{equation}
so the retrieval efficiency is
\begin{equation}
    \eta_{\mathrm{ret}}
    =
    2\Gamma_{c0}\int_0^\infty n(\tau)\,d\tau,
\end{equation}
which is exactly the same integral as in the constant-control case,
with $\Gamma_0$ and $g_0$ in place of $\Gamma_{\mathrm{eff}}$ and $g$.
Consequently, Eq.~\eqref{eq:retrieval} remains valid with
$\Gamma_{\mathrm{eff}}\to\Gamma_0$ and $g\to g_0$. In particular,
for $g_0=0$ the optimal retrieval efficiency
\begin{equation}
    \eta_{\mathrm{ret}}(g_0=0)
    =
    \frac{C}{1+C}
\end{equation}
is independent of the temporal shape of $\Omega_{se}(t)$, provided that the rates factorize through a common envelope $f(t)$ and the system remains in the stable regime.

Fig.~\ref{retrieval stable}(a) shows the retrieval efficiency in the stable regime with $g_{\mathrm{eff}}>\delta$ as a function of the ratio $\mu/\Gamma_{\mathrm{eff}}$. The efficiency increases monotonically with the effective coupling strength $g_{\mathrm{eff}}$ and can surpass unity, as indicated by the dotted line. This behavior signifies the presence of amplification effects originating from unwanted couplings. This trend becomes increasingly pronounced with higher cooperativity, which arises from an enhanced coupling strength. This sub-regime can emerge in systems characterized by nearly degenerate ground states or in Zeeman-type $\Lambda$ configurations in the limit of small detuning $\delta$. Fig.~\ref{retrieval stable}(b) corresponds to the regime $g_{\mathrm{eff}}<\delta$ and exhibits a qualitatively similar behavior; however, for a lower cooperativity, for example $C=1$, the retrieval efficiency predominantly remains below unity. This is the regime relevant to many warm-vapor memory experiments \cite{Hosseini2011,Pinel2013,Namazi2017,guo2019high,ma2022high,wu2025ai}. The fixed parameters employed in all plots are selected to ensure the validity of the adiabatic elimination of the optically excited states and the cavity mode, as well as the applicability of the bad-cavity approximation, while remaining representative of parameter regimes characteristic of warm-vapor-based quantum memories.

%In the absence of an initial spin excitation, the usual retrieval efficiency is not defined, since there is no signal to retrieve. Instead, the relevant quantity is the number of noise photons generated by the parametric term from vacuum fluctuations. In the stable regime,
%\begin{equation}
    %\Gamma_{\mathrm{eff}}^2+\delta^2-g_{\mathrm{eff}}^2>0,
%\end{equation}
%this number remains finite and is given by
%\begin{equation}
%N_{\mathrm{noise}}=\frac{C}{1+C}\,\frac{g_{\mathrm{eff}}^2}{\Gamma_{\mathrm{eff}}^2+\delta^2-g_{\mathrm{eff}}^2}.
%\label{eq:spurious photons}
%\end{equation}
%Therefore, the unwanted parametric coupling contributes a finite background of spurious output photons, even when no spin excitation is stored initially. This noise is suppressed when the detuning dominates, $\delta\gg g_{\mathrm{eff}}$, and increases as the system approaches the parametric threshold $g_{\mathrm{eff}}^2=\Gamma_{\mathrm{eff}}^2+\delta^2$.

\subsection{Retrieval efficiency in the threshold and unstable regimes}
\label{sec:detailed-ret-unstable}
We now move beyond the bounded regime and consider the finite-time behavior at and beyond threshold. The dynamics become unstable when the largest drift eigenvalue is positive, namely
\begin{equation}
\mu>\Gamma_{\mathrm{eff}}
\quad \Longleftrightarrow \quad
g^2_{\text{eff}}>\Gamma_{\mathrm{eff}}^2+\delta^2.
\label{eq:unstable-cond}
\end{equation}
In this regime, the population and output photon number grow exponentially with time, so an infinite-time retrieval ``efficiency'' is not well-defined. In this regime, the signal population becomes
\begin{equation}
n(t)=\frac{e^{-2\Gamma_{\mathrm{eff}} t}}{2\mu^2}\Big[3g^2_{\text{eff}}\cosh(2\mu t)-\bigl(g^2_{\text{eff}}+2\delta^2\bigr)\Big].
\label{eq:n-unst}
\end{equation}

The photon flux into the desired output mode is
\begin{equation}
    \Phi_{\mathrm{sig}}(t)=2\Gamma_c\,n(t),
\end{equation}
so we can define the finite-time retrieval efficiency (per initial excitation)
with a cut-off time $T$ as
\begin{equation}
    \eta_{\mathrm{ret}}(T)
    \equiv
    \int_0^T \Phi_{\mathrm{sig}}(t)\,dt
    =
    2\Gamma_c\int_0^T n(t)\,dt.
    \label{eq:eta-cut-def}
\end{equation}
Using Eq.~(\ref{eq:n-unst}) and evaluating the integrals gives
\begin{widetext}
\begin{equation}
\eta_{\mathrm{ret}}(T)=
\frac{C}{1+C}\frac{\Gamma_{\mathrm{eff}}}{\mu^2}\Bigg\{\frac{3g^2_{\text{eff}}}{4}\left[\frac{e^{2(\mu-\Gamma_{\mathrm{eff}})T}-1}{\mu-\Gamma_{\mathrm{eff}}}+\frac{1-e^{-2(\Gamma_{\mathrm{eff}}+\mu)T}}{\Gamma_{\mathrm{eff}}+\mu}\right]
-\frac{g^2_{\text{eff}}+2\delta^2}{2\Gamma_{\mathrm{eff}}}\bigl(1-e^{-2\Gamma_{\mathrm{eff}}T}\bigr)\Bigg\}
\label{eq:eta-cut-final}
\end{equation}
\end{widetext}
Eq.~(\ref{eq:eta-cut-final}) is finite for any fixed $T$ and applies in particular in the unstable regime. For $T\gg 1/\Gamma_{\text{eff}}$, the leading behavior is given by
\begin{equation}
    \eta_{\mathrm{ret}}(T)
    \sim
    \frac{C}{1+C}\,
    \frac{3g^2_{\text{eff}}\,\Gamma_{\mathrm{eff}}}{4\mu^2(\mu-\Gamma_{\mathrm{eff}})}
    \,e^{2(\mu-\Gamma_{\mathrm{eff}})T},
\end{equation}
reflecting exponential gain. In practice, pump depletion and
nonlinearities clamp this growth, so Eq.~(\ref{eq:eta-cut-final}) should be interpreted as linear short-time prediction.

The stability boundary is reached at
\begin{equation}
    g^2_{\text{eff}}=\Gamma_{\mathrm{eff}}^2+\delta^2
    \qquad\Longleftrightarrow\qquad
    \mu=\Gamma_{\mathrm{eff}}.
    \label{eq:threshold-cond}
\end{equation}
At this critical point, the largest drift eigenvalue vanishes, and the
linear dynamics becomes marginal: the infinite-time retrieval integral
does not converge, so one must also keep a finite cut-off time $T$.

Starting from the finite-time efficiency in the general case,
Eq.~(\ref{eq:eta-cut-final}), we take the limit $\mu\to\Gamma_{\mathrm{eff}}$. Using
\begin{equation}
    \lim_{\mu\to\Gamma_{\mathrm{eff}}}
    \frac{e^{2(\mu-\Gamma_{\mathrm{eff}})T}-1}{\mu-\Gamma_{\mathrm{eff}}}=2T,
\end{equation}
and $\Gamma_{\mathrm{eff}}+\mu \to 2\Gamma_{\mathrm{eff}}$, we obtain the threshold (critical) finite-time retrieval efficiency
\begin{widetext}   
\begin{equation}
\eta_{\mathrm{ret}}(T)=\frac{C}{1+C}\frac{1}{\Gamma_{\mathrm{eff}}}\Bigg\{\frac{3g^2_{\text{eff}}}{2}T+\frac{3g^2_{\text{eff}}}{8\Gamma_{\mathrm{eff}}}\bigl(1-e^{-4\Gamma_{\mathrm{eff}}T}\bigr)-\frac{g^2_{\text{eff}}+2\delta^2}{2\Gamma_{\mathrm{eff}}}\bigl(1-e^{-2\Gamma_{\mathrm{eff}}T}\bigr)\Bigg\}.
\label{eq:eta-threshold}
\end{equation}
\end{widetext}

For a long time $T\gg 1/\Gamma_{\mathrm{eff}}$, the exponentials vanish and Eq.~(\ref{eq:eta-threshold}) grows linearly with the time $T$:
\begin{equation}
\eta_{\mathrm{ret}}(T)\sim\frac{C}{1+C}\left(\frac{3g^2_{\text{eff}}}{2\Gamma_{\mathrm{eff}}}\right) T.
\end{equation}
In contrast to the stable regime, characterized by a finite asymptotic limit, and the unstable regime, marked by exponential growth, this linear divergence signifies marginal amplification precisely at the threshold.

Fig.~\ref{retrieval unstable} shows the retrieval efficiency as a function of $\Gamma_{\mathrm{eff}}T$ for the threshold and unstable regimes. For sufficiently long times, the unstable curves overtake the threshold curves, since the former grow exponentially, whereas the latter increase only linearly. Increasing the cooperativity shifts both sets of curves upward, showing that stronger light–matter coupling enhances the output in both regimes. As the efficiency is not bounded by the evolution time and far exceeds unity, the system effectively behaves as a photon source rather than a quantum memory.

\section{Storage Dynamics}
\label{sec:storage}

Unlike retrieval, the write process is driven directly by both the desired signal channel and the unwanted signal-induced channel, so the reduced dynamics is not related to the read process by simple time reversal. The derivation below starts from the exact finite-time storage kernel and then extracts the stable, threshold, and unstable limits. This construction also makes explicit why the write and read efficiencies no longer coincide once the unwanted couplings are present.

\subsection{Storage efficiency in the stable regime}
\label{sec:detailed-storage}

We now derive the storage formulas quoted above from the effective write dynamics
\begin{equation}
    \dot{\hat b}(t)
    =
    -\Gamma_{\mathrm{eff}}\hat b
    +g_{\mathrm{eff}}e^{-2i\delta t}\hat b^\dagger
    -J_1\hat a_{\rm in}
    -J_2e^{-2i\delta t}\hat a_{\rm in}^\dagger,
    \label{eq:b-storage-general}
\end{equation}
where $J_1=\sqrt{2\kappa N}\Omega^*_{se}G_{ge}/\alpha$ is the desired signal coupling strength, and $J_2=\sqrt{2\kappa N}\gamma G^*_{se'}\Omega_{ge'}/[(\gamma+i\Delta)\alpha]$ is the unwanted signal coupling strength.

For a single-photon input pulse
\begin{equation}
    |1_\xi\rangle
    =
    \int_0^T dt\,\xi(t)\hat a_{\rm in}^\dagger(t)|0\rangle,
    \qquad
    \int_0^T dt\,|\xi(t)|^2=1,
\end{equation}
the storage amplitude into the collective mode is
\begin{equation}
    \beta_\xi(T)=\langle 0|\hat b(T)|1_\xi\rangle,
\end{equation}
and the corresponding signal storage efficiency is
\begin{equation}
    \eta_s[\xi]=|\beta_\xi(T)|^2.
\end{equation}

\begin{figure}
\centering
\includegraphics[scale=0.25]{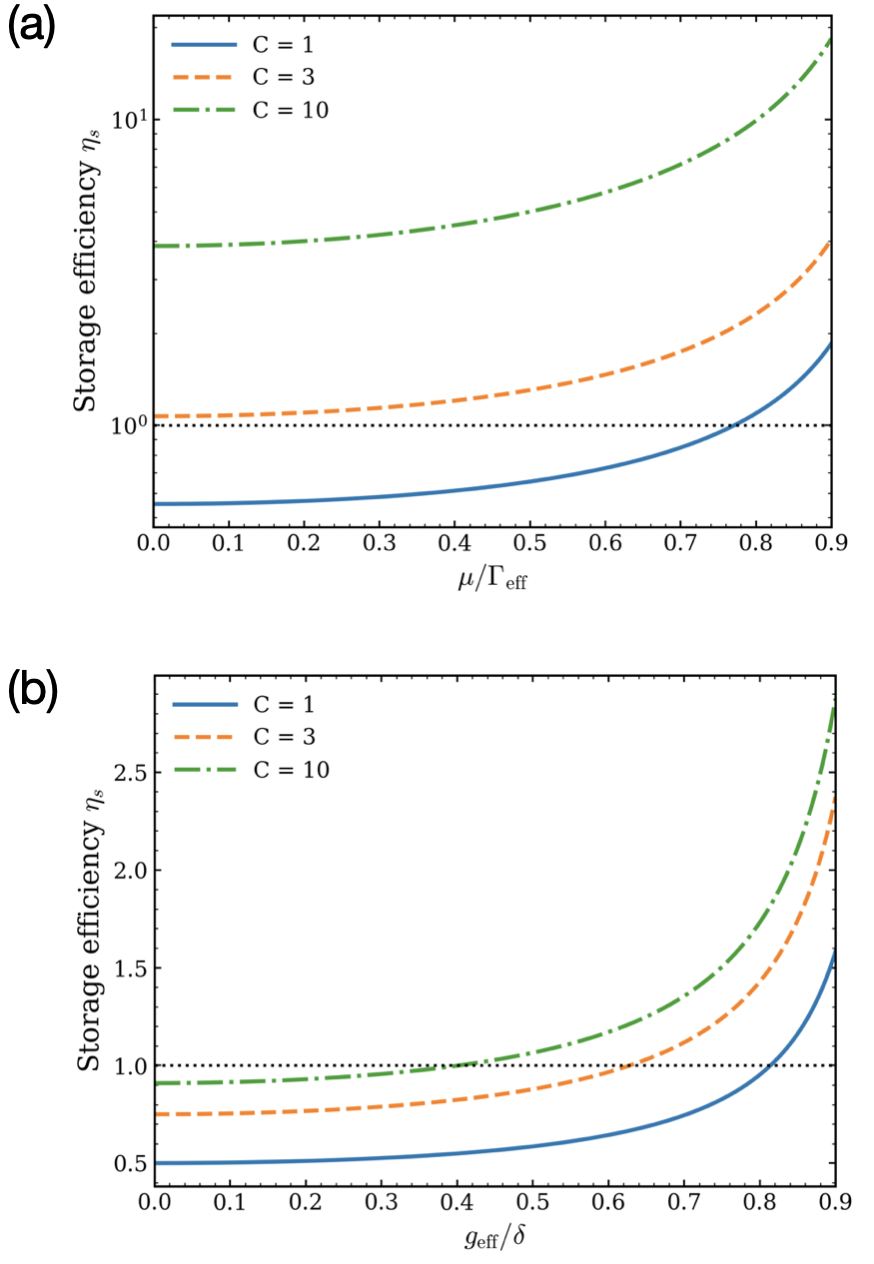}
\caption{
Storage efficiency \(\eta_s\) in (a) and (b), plotted for the stable regime
with \(g_{\mathrm{eff}}>\delta\) and \(g_{\mathrm{eff}}<\delta\), respectively. For (a) and (b), the horizontal axis is the dimensionless ratios
\(\mu/\Gamma_{\mathrm{eff}}\) and \(g_{\mathrm{eff}}/\delta\), respectively. For the panel (b), since in this case $\delta\gg\Gamma_{\mathrm{eff}}$, the performance is dictated by the ratio of $g_{\mathrm{eff}}$ to $\delta$. The curves correspond to cooperativities
\(C=|G_{ge}|^2N/(\gamma\kappa)=1,3,10\). The fixed parameters are the same as those in Fig.~\ref{retrieval stable}.
The horizontal dotted line in both panels marks \(\eta_{\mathrm{ret}}=1\), above which
the total output exceeds the ideal single-photon limit and indicates amplification.
}
\label{storage stable}
\end{figure}

For analytical transparency, we specialize to the case in which
$\Gamma_{\mathrm{eff}}, J_1, g_{\mathrm{eff}}, J_2$ are constant during the storage process. This regime already captures the central physics relevant to the optimal efficiency, namely the competition between the desired signal channel, the effective damping, and the amplification channel. Allowing these coefficients to vary in time does not change the structure of the optimization problem, since the control field simply induces time-dependent kernels, but it replaces the closed-form solution by a time-ordered evolution that must in general be treated numerically. The constant-coefficient case therefore provides the most transparent analytical benchmark. 

In the rotating-frame 

\begin{equation}
    \hat c(t)=e^{i\delta t}\hat b(t),
\end{equation}
the dynamics is governed by the drift matrix with eigenvalues
\begin{equation}
    \lambda_\pm=-\Gamma_{\mathrm{eff}}\pm\mu,
    \qquad
    \mu\equiv\sqrt{g_{\mathrm{eff}}^2-\delta^2}.
\end{equation}

Again, the stable regime separates into two sub-regimes:
\begin{align}
&\text{(i)}\quad g_{\mathrm{eff}}\ge \delta,
\quad
g_{\mathrm{eff}}^{2}<\Gamma_{\mathrm{eff}}^{2}+\delta^{2},
\label{eq:storestable-realmu}\\
&\text{(ii)}\quad g_{\mathrm{eff}}<\delta.
\label{eq:storestable-imagmu}
\end{align}
In sub-regime (i), \(\mu=\sqrt{g_{\mathrm{eff}}^{2}-\delta^{2}}\) is real and satisfies
\(0\le \mu<\Gamma_{\mathrm{eff}}\). In sub-regime (ii), \(\mu\) is purely imaginary,
\(\mu=i\mu'\) with \(\mu'=\sqrt{\delta^{2}-g_{\mathrm{eff}}^{2}}\), and the system is always stable. In the following, we first focus on the first sub-regime. 

The signal kernel can then be written as
\begin{equation}
    f(u)=A_+e^{s_+u}+A_-e^{s_-u},
    \qquad
    u= T-t,
\end{equation}
with
\begin{equation}
    A_\pm
    =
    \frac{1}{2}
    \left[
        J_1\left(1\pm\frac{i\delta}{\mu}\right)
        \pm
        \frac{g_{\mathrm{eff}}J_2^*}{\mu}
    \right],\quad s_\pm=\lambda_\pm-i\delta.
    \label{eq:Apm-compact}
\end{equation}
Hence
\begin{equation}
    \beta_\xi(T)=\int_0^T dt\,f(T-t)\xi(t),
\end{equation}
so that
\begin{equation}
    \eta_s[\xi]
    =
    \left|
        \int_0^T dt\,f(T-t)\xi(t)
    \right|^2.
\end{equation}
The optimal input mode is
\begin{equation}
    \xi_{\rm opt}(t)\propto f^*(T-t),
\end{equation}
and the corresponding optimal signal storage efficiency is simply
\begin{equation}
    \eta_s(T)=\int_0^T du\,|f(u)|^2.
\end{equation}
Substituting the form of \(f(u)\) gives the exact finite-time result
\begin{equation}
    \eta_s(T)
    =
    \sum_{\alpha,\beta=\pm}
    A_\alpha A_\beta^*
    \frac{e^{(s_\alpha+s_\beta^*)T}-1}{s_\alpha+s_\beta^*}.
    \label{eq:eta-opt-exact-compact}
\end{equation}
This can be further written as 
\begin{widetext}
\begin{equation}
    \eta_s(T)=
    \frac{|A_+|^2\!\left(1-e^{-2(\Gamma_{\mathrm{eff}}-\mu)T}\right)}{2(\Gamma_{\mathrm{eff}}-\mu)}
    +
    \frac{|A_-|^2\!\left(1-e^{-2(\Gamma_{\mathrm{eff}}+\mu)T}\right)}{2(\Gamma_{\mathrm{eff}}+\mu)}+
    \Re\left[
        A_+A_-^*
        \frac{1-e^{-2\Gamma_{\mathrm{eff}}T}}{\Gamma_{\mathrm{eff}}}
    \right].
\label{eq:eta-opt-finiteT-compact}
\end{equation}
\end{widetext}

We note that the desired signal coupling satisfies the following condition.
\begin{equation}
    \frac{|J_1|^2}{2}=\Gamma_c,
    \qquad
    \Gamma_c=\frac{C}{C+1}\Gamma_{\mathrm{eff}},
\end{equation}
where \(C\) is the cavity cooperativity.

In the limit $T\gg\Gamma_{\mathrm{eff}}$, we obtain
\begin{equation}
    \eta_s
    =
    \frac{|A_+|^2}{2(\Gamma_{\mathrm{eff}}-\mu)}
    +
    \frac{|A_-|^2}{2(\Gamma_{\mathrm{eff}}+\mu)}
    +
    \Re\!\left[
        \frac{A_+A_-^*}{\Gamma_{\mathrm{eff}}}
    \right].
    \label{eq:eta-opt-stable-final}
\end{equation}
We can further simplify it as follows
\begin{equation}
    \eta_s
    =
    \frac{C}{C+1}
    \left[
        \frac{|B_+|^2}{4(1-r)}
        +
        \frac{|B_-|^2}{4(1+r)}
        +
        \frac{1}{2}\Re(B_+B_-^*)
    \right],
    \label{eq:storage}
\end{equation}
where $r=\mu/\Gamma_{\mathrm{eff}}\in[0,1)$, and 
\begin{equation}
    B_\pm=
    1\pm\frac{i\delta}{\mu}\pm\frac{g_{\mathrm{eff}}}{\mu}\tilde J_2^*,
\end{equation}
and $\tilde J_2\equiv J_2/J_1$. $A_\pm$ and $B_\pm$ are related by $A_\pm=J_1 B_\pm/2$. Thus, in the stable regime, the final storage efficiency is the optimal value in the noiseless case $C/(C+1)$ multiplied by a correction factor determined by the noise amplitudes $g_{\mathrm{eff}}$ and $J_2$ and the detuning $\delta$. When we are in the second sub-regime, i.e., $\delta>g_{\mathrm{eff}}$, the storage efficiency becomes

\begin{equation}
    \eta_s
    =
    \frac{C}{C+1}
    \left[
        \frac{|B'_+|^2}{4}
        +
        \frac{|B'_-|^2}{4}
        +
        \frac{1}{2}\Re\!\left(\frac{B'_+B'^*_-}{1-ir'}\right)
    \right],
    \label{eq:imstorage}
\end{equation}
where $r'=\mu'/\Gamma_{\mathrm{eff}}$ with $\mu'=\sqrt{\delta^2-g^2_{\mathrm{eff}}}$ and
\begin{equation}
    B'_\pm=
    1\pm\frac{\delta}{\mu'}\mp\frac{ig_{\mathrm{eff}}}{\mu'}\tilde J_2^*.
\end{equation}

\begin{figure}
\centering
\includegraphics[scale=0.45]{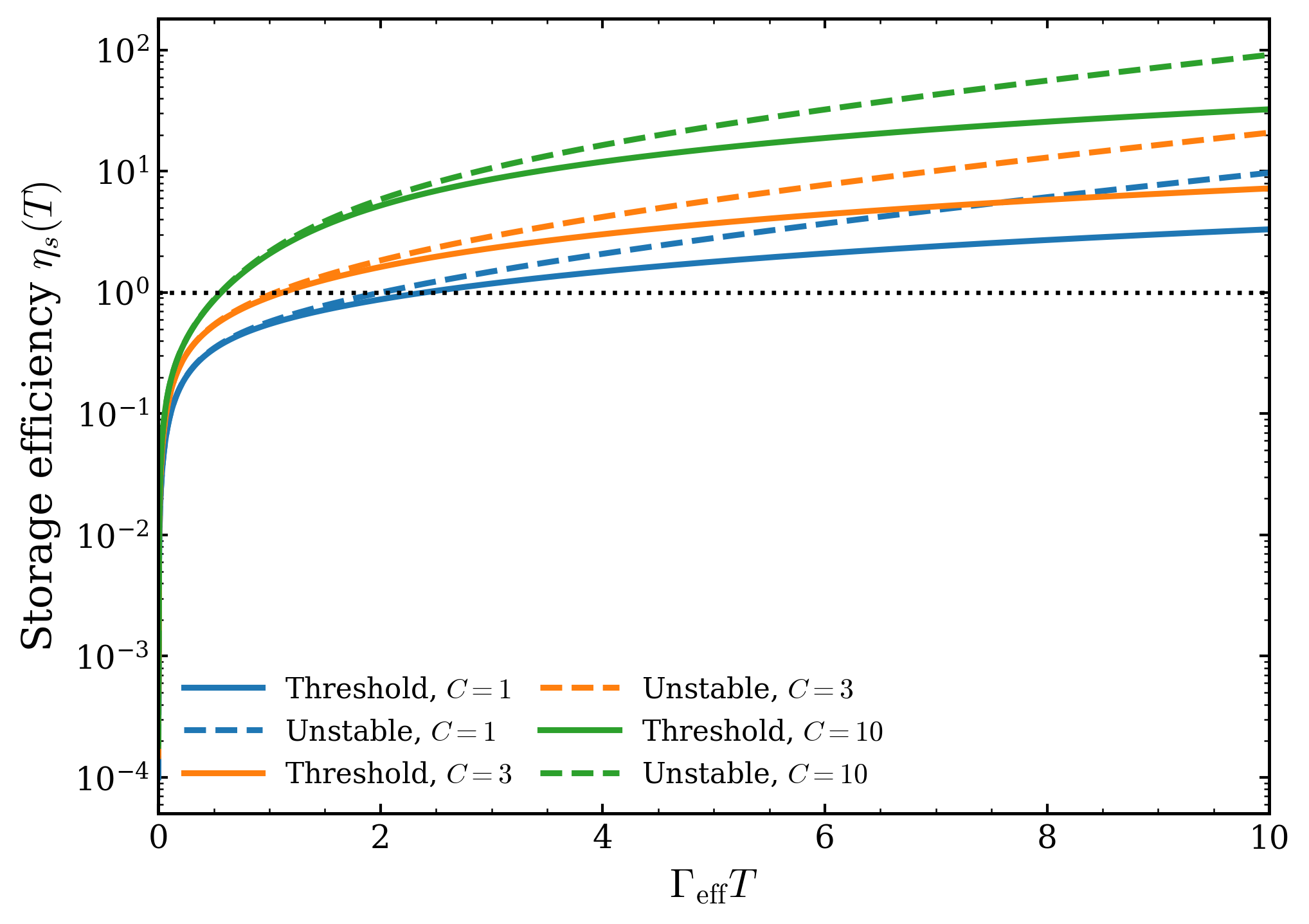}
\caption{
Storage efficiency for the threshold and unstable regimes, plotted as a function of the dimensionless total protocol time $\Gamma_{\mathrm{eff}}T$. The threshold regime satisfies $\Gamma_{\mathrm{eff}}=\mu$, and the unstable regime satisfies $\Gamma_{\mathrm{eff}}<\mu$. The curves correspond to cooperativities
\(C=|G_{ge}|^2N/(\gamma\kappa)=1,3,10\) in each regime. The fixed parameters are the same as those in Fig.~\ref{retrieval unstable} with $\delta=1$ MHz. The horizontal dotted line marks \(\eta_{\mathrm{tot}}=1\), above which
the total output exceeds the ideal single-photon limit and indicates amplification. In these two cases, the memory effectively operates as a photon source, emitting photons continuously over time.}
\label{storage unstable}
\end{figure}

Fig.~\ref{storage stable}(a) shows the storage efficiency in the stable regime with $g_{\mathrm{eff}}>\delta$ as a function of the ratio $\mu/\Gamma_{\mathrm{eff}}$. The storage efficiency also increases monotonically with the effective coupling strength $g_{\mathrm{eff}}$ and can surpass unity as a result of amplification originating from unwanted couplings. With higher cooperativity, the amplification increases, arising from an enhanced coupling strength. This sub-regime here also corresponds to the physical systems with nearly degenerate ground states or Zeeman-type $\Lambda$ configurations in the limit of small detuning $\delta$. Fig.~\ref{storage stable}(b) corresponds to the regime $g_{\mathrm{eff}}<\delta$ and exhibits a qualitatively similar behavior, but for a lower cooperativity, for example, $C=1$, the storage efficiency predominantly remains below unity. Although the storage efficiency in the stable regime does not coincide with the retrieval efficiency as a consequence of broken time-reversal symmetry, the overall behavior of the two quantities exhibits a similar trend. Again, this regime pertains to many warm-vapor memory experiments, and the fixed parameters used in Fig.~\ref{storage stable} and Fig.~\ref{storage unstable} are chosen so that adiabatic elimination and bad-cavity approximation are ensured.

\subsection{Storage efficiency in the threshold and unstable regimes}

At the threshold,
\begin{equation}
    \Gamma_{\mathrm{eff}}=\mu
    \quad
    \Longleftrightarrow
    \quad
    g_{\mathrm{eff}}^2=\Gamma_{\mathrm{eff}}^2+\delta^2,
\end{equation}
the dominant mode becomes marginal. When $\mu=\Gamma_{\mathrm{eff}}$,
\begin{equation}
\begin{aligned}
    \eta_s(T)
    &=
    |A_+|^2T
    +
    \frac{|A_-|^2}{4\Gamma_{\mathrm{eff}}}\left(1-e^{-4\Gamma_{\mathrm{eff}}T}\right)
    \\
    &+
    \frac{\Re(A_+A_-^*)}{\Gamma_{\mathrm{eff}}}\left(1-e^{-2\Gamma_{\mathrm{eff}}T}\right).
\end{aligned}
\end{equation}
We can further express it in terms of $B_\pm$ in the limit of $T\gg1/\Gamma_{\mathrm{eff}}$
\begin{equation}
\eta_s(T)=\frac{C}{C+1}\frac{|B_+|^2}{2}\Gamma_{\mathrm{eff}}T.  
\end{equation}
Consequently, at the threshold, the optimized signal efficiency exhibits a linear dependence on the parameter $T$.

If the following conditions are satisfied
\begin{equation}
    \Gamma_{\mathrm{eff}}<\mu
    \quad
    \Longleftrightarrow
    \quad
    g_{\mathrm{eff}}^2>\Gamma_{\mathrm{eff}}^2+\delta^2,
\end{equation}
the exact finite-time result remains Eq.~\eqref{eq:eta-opt-finiteT-compact}, but for real \(\mu>\Gamma_{\mathrm{eff}}\) and the long time limit $T\gg1/\Gamma_{\mathrm{eff}}$, we have

\begin{equation}
    \eta_s(T)
    \sim
    \frac{C}{C+1}\,
    \frac{|B_+|^2}{4(r-1)}
    e^{2(r-1)\Gamma_{\mathrm{eff}}T}.
\end{equation}
In this regime, the system no longer behaves as a memory, but rather as an amplifier.

We could note that when \(g_{\mathrm{eff}}=J_2=0\), one has $B_+=2$ and $B_-=0$ as $\mu=i\delta$. Since $\mu$ is an imaginary number, the storage efficiency in the stable regime is given by Eq.~\eqref{eq:imstorage}. Thus, we end up having

\begin{equation}
\eta_s=\frac{C}{C+1}.    
\end{equation}
Thus, the well-known result in the noiseless case is recovered. In the limit $g_{\mathrm{eff}}=J_2=0$, the dynamics is number-conserving, and optimal storage and retrieval are related by time reversal, yielding identical efficiencies. By contrast, when the terms $g_{\mathrm{eff}}$ and $J_2$ are present, the reduced dynamics becomes Bogoliubov rather than passive, mixing annihilation and creation operators. The control field can then inject energy into the system, so the write and read processes are no longer related by simple time reversal, and the optimal storage and retrieval efficiencies generally differ.

Fig.~\ref{storage unstable} shows the storage efficiency as a function of $\Gamma_{\mathrm{eff}}T$ for the threshold and unstable regimes. Like the retrieval efficiency in these two regimes, the unstable curves eventually surpass the threshold curves at sufficiently long times because the former grow exponentially, whereas the latter increase only linearly. Increasing the cooperativity shifts both sets of curves upward, indicating that stronger light–matter coupling enhances the output in both regimes. Since the storage efficiency is unbounded in time and can greatly exceed unity, the write process no longer represents faithful storage of the input photon. Instead, the unwanted couplings generate additional collective spin excitations, so the system effectively behaves as a spin-wave amplifier rather than as a quantum memory. However, in a real ensemble, this growth is ultimately limited by the finite number of atoms and by nonlinear saturation beyond the low-excitation approximation, which is beyond the scope of this work.

\section{Memory Performance and Regime Comparison}
\label{sec:fidelity}

With the storage and retrieval efficiencies in hand, we can quantify the memory performance in the stable regime. A commonly used operational benchmark infers the fidelity from the photons retrieved in the absence of both an input field and any initial spin excitation \cite{nunn2017theory,asadiunwanted}. In the present problem, however, that benchmark does not capture the full effect of unwanted couplings, because noise can also be generated during storage. We therefore define the stable-regime fidelity as the fraction of retrieved photons that belong to the desired signal,
\begin{equation}
F=\frac{\mathrm{SNR}}{1+\mathrm{SNR}}=\frac{N_{\mathrm{sig}}}{N_{\mathrm{tot}}},
\end{equation}
where \(N_{\mathrm{sig}}\) is the desired retrieved signal and \(N_{\mathrm{tot}}\) is the total retrieved output. Using the noiseless memory as the benchmark, we take
\begin{equation}
N_{\mathrm{sig}}=\left(\frac{C}{C+1}\right)^2,
\qquad
N_{\mathrm{tot}}=\eta_s\eta_{\mathrm{ret}},
\end{equation}
where $\eta_{\mathrm{ret}}$ is given by Eqs.~\eqref{eq:retrieval} and $\eta_s$ is given by Eq.~\eqref{eq:storage} and  Eq.~\eqref{eq:imstorage}, respectively for th two sub-regimes in the stable regime. The operational fidelity then becomes
\begin{equation}
F=\frac{\left(\frac{C}{C+1}\right)^2}{\eta_s\eta_{\mathrm{ret}}}.
\label{eq:fidelity}
\end{equation}
We can now compare the efficiency--fidelity trade-off in the two stable sub-regimes and then contrast them with the threshold and unstable cases. 

\begin{figure}
\centering
\includegraphics[scale=0.25]{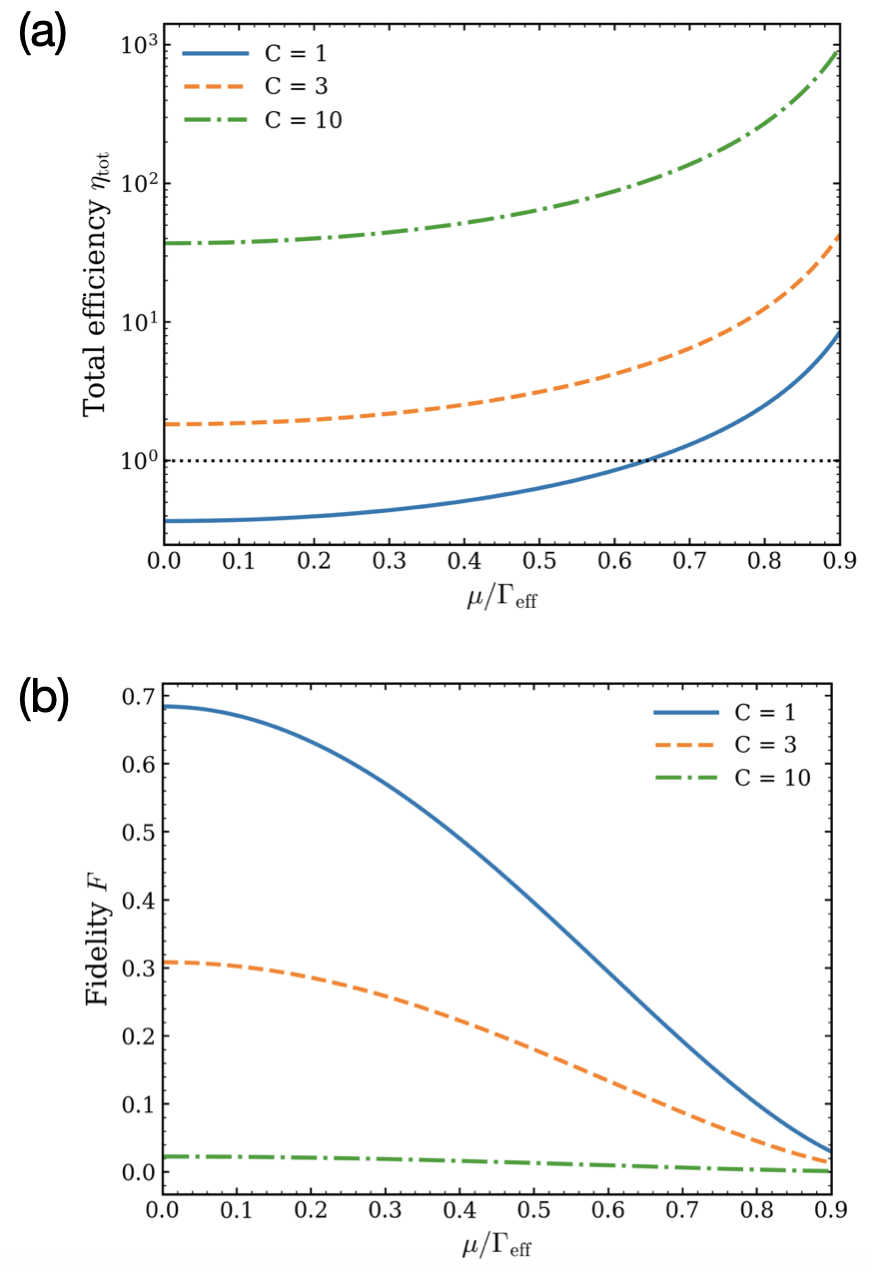}
\caption{
Total efficiency \(\eta_{\mathrm{tot}}\) in (a) and fidelity
\(F\) in (b), plotted for the stable regime
with $\Gamma_{\mathrm{eff}}>\mu$ and \(g_{\mathrm{eff}}>\delta\). The curves correspond to cooperativities
\(C=|G_{ge}|^2N/(\gamma\kappa)=1,3,10\). 
The parameters used here are chosen to be the same as the ones used in Fig~\ref{retrieval stable}(a). The horizontal dotted line in panel (a) marks \(\eta_{\mathrm{tot}}=1\), above which
the total output exceeds the ideal single-photon limit and indicates amplification.
}
\label{ge plot}
\end{figure}

\begin{figure}
\centering
\includegraphics[scale=0.25]{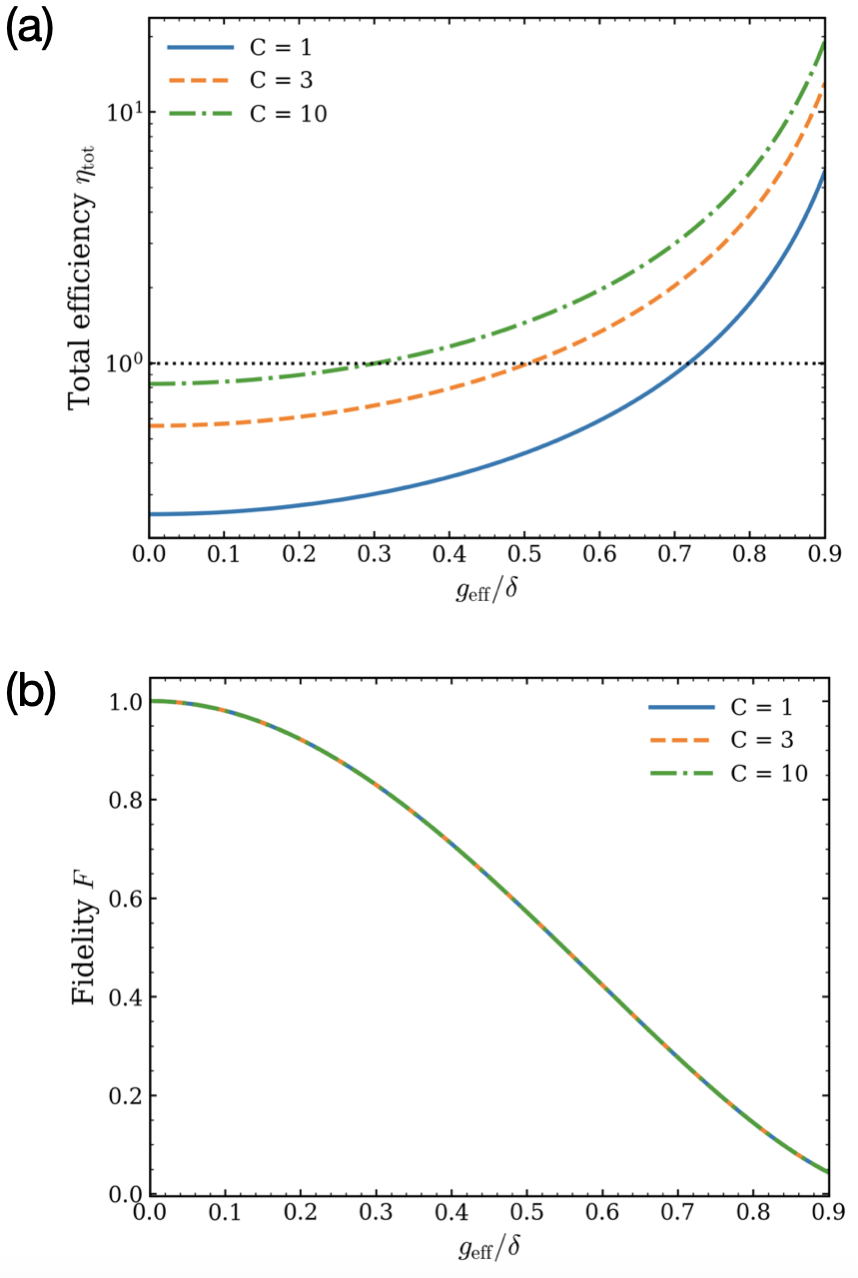}
\caption{
Total efficiency \(\eta_{\mathrm{tot}}\) in (a) and fidelity
\(F\) in (b), plotted as functions of the normalized effective unwanted coupling \(g_{\mathrm{eff}}/\delta\) in the stable regime. The curves correspond to cooperativities
\(C=|G_{ge}|^2N/(\gamma\kappa)=1,3,10\).
The fixed parameters are the same as those in Fig.~\ref{ge plot} but with a different ground-state splitting \(\delta=6.8~\mathrm{GHz}\). The horizontal dotted line in panel (a) marks \(\eta_{\mathrm{tot}}=1\), above which
the total output exceeds the ideal single-photon limit and indicates amplification. This is the case relevant to many warm-vapor memory experiments \cite{Hosseini2011,Pinel2013,Namazi2017,guo2019high,ma2022high,wu2025ai}.}
\label{rb plot}
\end{figure}

\begin{figure}
\centering
\includegraphics[scale=0.44]{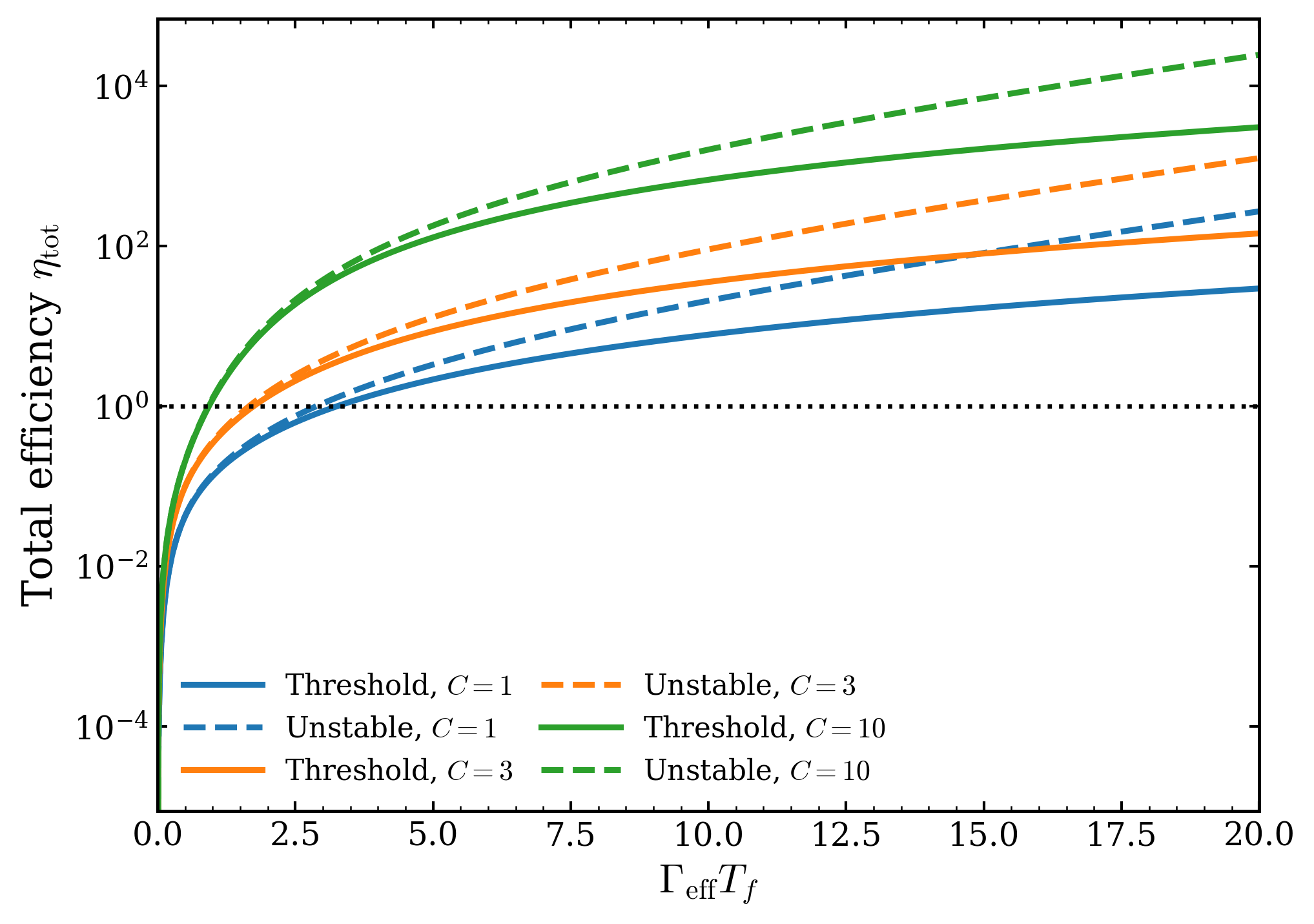}
\caption{
Total efficiency for the threshold and unstable regimes, plotted as a function of the dimensionless total protocol time $\Gamma_{\mathrm{eff}}T_f$, where $T_f=2T$ assumes equal storage and retrieval durations. The threshold regime satisfies $\Gamma_{\mathrm{eff}}=\mu$, and the unstable regime satisfies $\Gamma_{\mathrm{eff}}<\mu$. The curves correspond to cooperativities
\(C=|G_{ge}|^2N/(\gamma\kappa)=1,3,10\) in each regime. The fixed parameters are the same as those in Fig.~\ref{ge plot}. The horizontal dotted line marks \(\eta_{\mathrm{tot}}=1\), above which the total output exceeds the ideal single-photon limit and indicates amplification. In these two cases, the fidelity is no longer a meaningful figure of merit, since the memory effectively operates as a photon source, emitting photons continuously over time.}
\label{threun plot}
\end{figure}

Figure~\ref{ge plot} shows the total efficiency and fidelity for the case $g_{\mathrm{eff}}>\delta$, plotted against the dimensionless ratio $\mu/\Gamma_{\mathrm{eff}}$. For the parameter set used in Fig.~\ref{ge plot}(a), the total efficiency rises rapidly with increasing $\mu/\Gamma_{\mathrm{eff}}$ and can cross the reference value $\eta_{\mathrm{tot}}=1$, marked by the horizontal dotted line. At the same time, the fidelity in Fig.~\ref{ge plot}(b) decreases monotonically. The figure makes clear that a larger total output does not necessarily imply better memory performance; instead, it can reflect the increasing role of gain and unwanted noise. This trend becomes more pronounced as the cooperativity $C$ increases: a larger $C$ enhances the total output but also pushes the system farther away from the ideal single-photon memory regime.

Figure~\ref{rb plot} shows the total efficiency and fidelity for the complementary stable regime $\delta>g_{\mathrm{eff}}$, using the normalized coupling $g_{\mathrm{eff}}/\delta$ as the horizontal axis. In this case, the total efficiency again increases monotonically with the unwanted coupling strength. It can exceed unity for sufficiently large cooperativity, indicating the onset of amplification rather than genuine memory operation. By contrast, for smaller cooperativity, such as $C=1$, the total efficiency remains below unity over most of the plotted range. Meanwhile, the fidelity decreases steadily as $g_{\mathrm{eff}}/\delta$ increases, showing once more that stronger output does not imply higher memory quality. In the weak-coupling limit $g_{\mathrm{eff}}\ll\delta$, however, the fidelity remains close to unity, indicating that the memory can still operate with high quality in this regime.

The different cooperativity dependence of the fidelity in Figs.~\ref{ge plot}(b) and \ref{rb plot}(b) follows directly from the definition in Eq.~\eqref{eq:fidelity}. In the regime $\delta>g_{\mathrm{eff}}$, let $x=g_{\mathrm{eff}}/\delta$. Then both $\delta/\mu'$ and $g_{\mathrm{eff}}/\mu'$ depend only on $x$, with $\mu'=\sqrt{\delta^2-g_{\mathrm{eff}}^2}$. For the parameter set of Fig.~\ref{rb plot}, one has $\delta\gg\Gamma_{\mathrm{eff}}$ for all plotted cooperativities, so that $r'=\mu'/\Gamma_{\mathrm{eff}}\gg 1$. In this limit, the interference term in $\eta_s$ is strongly suppressed, and both $\eta_s$ and $\eta_{\mathrm{ret}}$ are approximately proportional to $C/(C+1)$ times a function of $x$ only. As a result,
\begin{equation}
    \eta_{\mathrm{ret}}\eta_s \approx \left(\frac{C}{C+1}\right)^2 S(x),
\end{equation}
so the explicit $C$ dependence largely cancels in $F$. This explains the near overlap of the fidelity curves in Fig.~\ref{rb plot}(b). The physical intuition is that, for \(\delta>g_{\mathrm{eff}}\) with \(\delta\gg\Gamma_{\mathrm{eff}}\), the detuning suppresses the coherent buildup of the unwanted gain channel, leaving \(g_{\mathrm{eff}}/\delta\) as the dominant parameter controlling the noise fraction. The cooperativity mainly changes the overall extraction efficiency into the cavity mode, so it enhances the desired signal and the unwanted output by almost the same multiplicative factor. This common factor cancels in the ratio defining \(F\), explaining why the fidelity curves in Fig.~\ref{rb plot}(b) nearly overlap even though the total efficiencies in Fig.~\ref{rb plot}(a) remain \(C\)-dependent.

By contrast, in the regime $g_{\mathrm{eff}}>\delta$, the horizontal variable is $x=\mu/\Gamma_{\mathrm{eff}}$, so at fixed $x$ one has
\begin{equation}
    g_{\mathrm{eff}}^2=x^2\Gamma_{\mathrm{eff}}^2+\delta^2.
\end{equation}
Since $\Gamma_{\mathrm{eff}}=|\Omega_{se}|^2/[\gamma(C+1)]$, changing $C$ also changes the ratio $\delta/\Gamma_{\mathrm{eff}}$. This ratio enters both the retrieval efficiency and the coefficients $B_\pm$ that determine the storage efficiency. Consequently, $\eta_{\mathrm{ret}}\eta_s$ is no longer simply proportional to $\left(C/(C+1)\right)^2$, the cancellation in $F$ is incomplete, and the fidelity acquires the pronounced cooperativity dependence seen in Fig.~\ref{ge plot}(b).

Figure~\ref{threun plot} finally compares the total efficiency in the threshold and unstable regimes as a function of the dimensionless total protocol time $\Gamma_{\mathrm{eff}}T_f$, where $T_f=2T$ assumes equal storage and retrieval durations. In the threshold regime, both the storage and retrieval efficiencies grow linearly with time, so the total efficiency increases algebraically with $T_f$. By contrast, in the unstable regime, both processes acquire exponential gain, leading to a much faster increase of $\eta_{\mathrm{tot}}$. The unstable curves overtake the threshold ones for sufficiently long times, as the unstable regime exhibits exponential growth compared to the linear growth in the threshold regime. Increasing the cooperativity shifts both sets of curves upward, showing that stronger light--matter coupling enhances the total output in both regimes. In these two cases, the fidelity is no longer a meaningful figure of merit, because the system effectively operates as a photon source rather than as a memory.

\section{Conclusion}
\label{sec:conclusion}
We have presented a fully quantum analytical study of a four-level cavity-based ensemble quantum memory in the presence of unwanted couplings from both the control and signal fields. Our treatment yields explicit expressions for the storage, retrieval, and fidelity efficiencies and identifies stable, threshold, and unstable dynamical regimes. The stable regime supports bounded memory dynamics, whereas the threshold and unstable regimes mark a crossover to amplification.

Our results show that high output efficiency does not by itself imply good memory performance: when unwanted couplings become sufficiently strong, $\eta_{\mathrm{tot}}>1$ may be accompanied by reduced fidelity, indicating gain rather than faithful storage and retrieval. In the noiseless limit, we recover the standard result $\eta=C/(1+C)$ and the usual time-reversal relation between optimal storage and retrieval in ideal cavity memories \cite{gorshkov2007photon}. Unwanted couplings break this symmetry and qualitatively modify the memory dynamics. For warm-vapor-relevant parameters, we find that the regime $\delta>g_{\mathrm{eff}}$, particularly $g_{\mathrm{eff}}\ll\delta$, remains compatible with high-fidelity operation, consistent with room-temperature and cavity-enhanced Raman-memory experiments \cite{reim2011single,saunders2016cavity,Hosseini2011,ma2022high,guo2019high,Namazi2017,wu2025ai}.

More broadly, our results complement earlier studies of four-wave-mixing-induced noise and fidelity degradation in absorptive memories \cite{lauk2013fidelity,nunn2017theory}, while extending them to a four-level setting with a full quantum treatment. Looking ahead, it will be interesting to apply this framework to emerging room-temperature memory platforms aimed at practical networking, including field-deployable warm-vapor memories, single-photon-compatible vapor-cell memories, synchronization and interference of independent room-temperature memories, and microfabricated vapor-cell implementations \cite{wang2022fielddeployable,buser2022singlephoton,davidson2023synchronization,gera2024hom,mottola2023optical}. In this sense, the present theory provides both a diagnostic criterion and a design tool for optimizing realistic absorptive quantum memories beyond idealized few-level models and for guiding their integration into scalable quantum-network architectures \cite{Wehnereaam9288,RevModPhys.95.045006}.

\section*{Acknowledgments}
This work is funded by NSERC Alliance quantum consortium grants ARAQNE and QUINT and the NRC High-throughput Secure Networks (HTSN) challenge program.

%\newpage

\appendix
\section{THE IMPACT OF NON-LINEAR TERMS}
\label{appa}

Here we justify the simplification from Eq.~\eqref{eq:adibatic} to Eq.~\eqref{eq:simp}. The nonlinear terms in Eq.~\eqref{eq:adibatic} can be neglected in the low-excitation regime because they are higher order in the small excitation amplitude. Since the derivation already assumes that almost all atoms remain in the ground state, we may introduce a bookkeeping parameter \(\varepsilon\) such that
\begin{equation}
\hat a_{\mathrm{in}}=O(\varepsilon),\quad
\hat\sigma_{gs}=O(\varepsilon).
\end{equation}
Because $\hat\beta=\sqrt{2\kappa}\,\gamma\,\hat a_{\mathrm{in}}-\Omega_{se}G_{ge}^*\hat\sigma_{gs}$, it follows that
\begin{equation}
\hat\beta \sim O(\varepsilon),\quad
\hat\beta^\dagger\hat\beta \sim O(\varepsilon^2).
\end{equation}
Hence the linear terms in Eq.~\eqref{eq:adibatic}, proportional to \(\hat\sigma_{gs}\), \(\hat\beta/\alpha\), and \(\hat\beta^\dagger/\alpha\), are all \(O(\varepsilon)\), whereas the nonlinear terms
\begin{equation}
\frac{\hat\beta^\dagger\hat\beta}{\alpha^2}\hat\sigma_{gs} \sim O(\varepsilon^3),
\quad
\frac{\hat\beta\hat\beta^\dagger}{\alpha^2}\hat\sigma_{gs}^\dagger \sim O(\varepsilon^3).
\end{equation}
They are therefore smaller by a factor \(O(\varepsilon^2)\), so Eq.~\eqref{eq:simp} is the leading-order weak-excitation truncation of Eq.~\eqref{eq:adibatic}. We emphasize that the linearized treatment is valid only while the amplification remains sufficiently weak. In the threshold and unstable regimes discussed in Sec.\ref{sec:retrieval} and Sec.\ref{sec:storage}, the linearized solution formally exhibits unbounded growth. In practice, however, pump depletion and higher-order nonlinearities eventually become important and saturate the dynamics. A quantitative treatment of this saturation is beyond the scope of the present work.

\bibliographystyle{apsrev4-1}
\bibliography{mybib}

\end{document}